\documentclass[preprint,12pt]{elsarticle}

\usepackage{amssymb}
\usepackage{amsmath}
\usepackage{atlasphysics}
\usepackage{subfigure}

\begin{document}

\begin{frontmatter}



\title{Measured Effectiveness of Deep N-well Substrate Isolation in a 65nm Pixel Readout Chip Prototype}

\author[label1]{Peilian Liu \footnote{Corresponding author, email: liupl@ihep.ac.cn}}
\author[label2]{Maurice Garcia-Sciveres}
\author[label2]{Timon Heim}
\author[label2]{Amanda Krieger}
\author[label2]{Dario Gnani}

\address[label1]{Institute of High Energy Physics, Beijing, 100049, CN}
\address[label2]{Physics Division, Lawrence Berkeley National Laboratory, Berkeley, CA 94720, USA}

\begin{abstract}
The same charge sensitive preamplifier and discriminator circuit with different isolation strategies has been tested to compare the isolation of both analog and digital circuits from the substrate of a 65\,nm bulk CMOS process to the isolation of only digital circuits, tying analog ground locally to the substrate. This study will show that the circuit with analog on the substrate and digital in deep N-well has better noise isolation between analog and digital.
\end{abstract}


\end{frontmatter}


\section{Introduction}
A significant concern for mixed signal circuits, and for detector front end Application Specific Integrated Circuits (ASICs) in particular, is the isolation of sensitive analog nodes from digital activity. Bulk CMOS technologies typically offer deep implants that can be used to electrically isolate individual transistors or full circuit blocks from the substrate. How to handle the substrate and what isolation strategy to use can depend on the particular CMOS process, but can be informed by general guidelines. It is normally not practical to prototype alternative isolation strategies in order to select one, as it requires to duplicate the same ASIC with alternative isolation in order to compare the two.

For this note we take advantage of a rare opportunity to prototype and test the same complex circuit with
different isolation strategies, and compare the results.
We compare the isolation of both analog and digital circuits from the substrate of a 65nm bulk CMOS process
(effectively leaving the substrate as a buffer with no electrical function)
to the isolation of only digital circuits, tying analog ground locally to the substrate.
We refer to the former as double isolation and to the latter as digital isolation.
A pixel readout matrix demonstrator ASIC, called FE65-P2~\cite{ref:FE65-P2}, was fabricated twice, once with double isolation
(original design from~\cite{ref:isolation}) and once with digital isolation, by simply removing the deep n-well
layer from the analog front end ``islands'' in pixel matrix with no other changes.

\begin{figure}[h]
  \centering
  \includegraphics[width=0.6\textwidth]{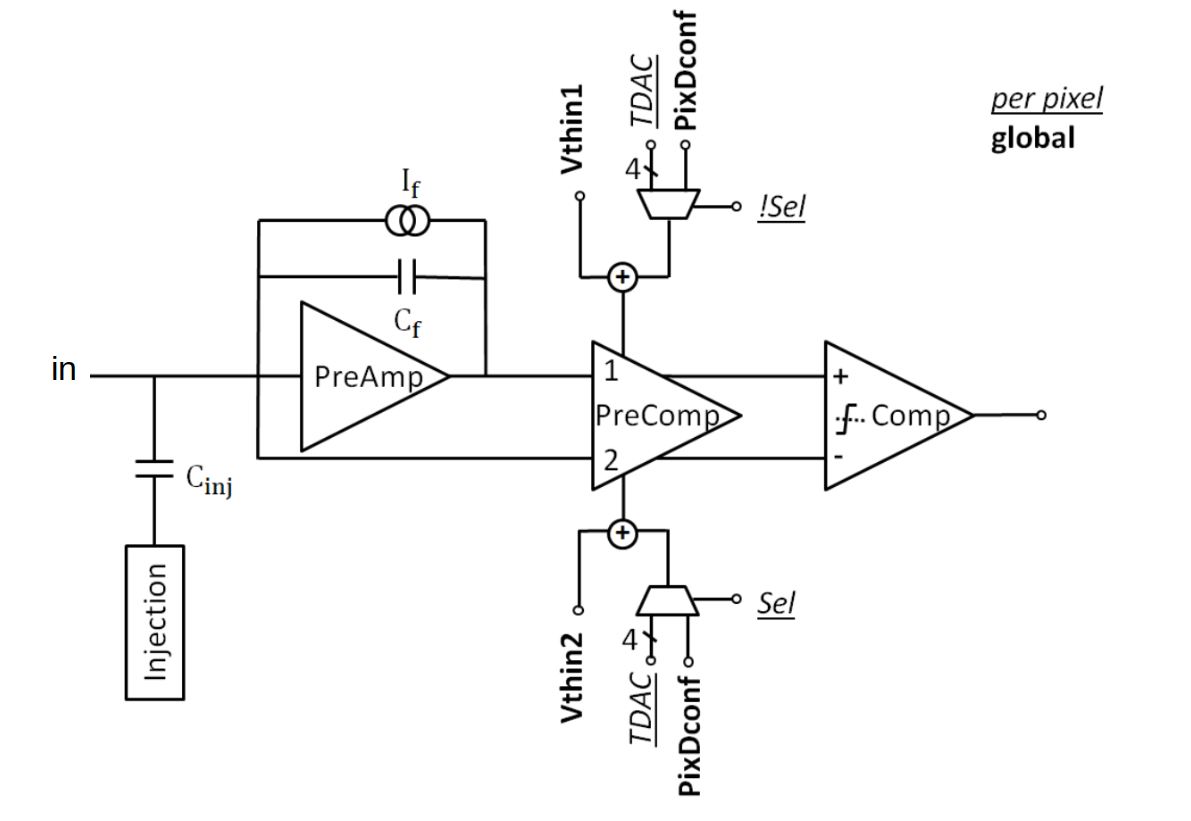}
  \caption{Schematic diagram of a pixel analog front end.}
\label{fig:analog}
\end{figure}

The FE65-P2 contains a matrix of 64 by 64 pixels on 50\,$\mu$m by 50\,$\mu$m pitch. Every pixel has a dedicated analog front end, consisting of a charge integrator, followed by a single ended to differential second stage feeding a differential comparator, shown in Fig.~\ref{fig:analog}.
The pixel threshold is controlled by two global, internal 8-bit DACs (Vthin1 and Vhin2). All pixels receive the same global threshold voltages, but each pixel's effective threshold will occur at a different DAC value due to transistor mismatch caused by production process variations. The FE65-P2 includes a 5-bit trim DAC (TDAC) in each pixel to compensate for this mismatch.

\begin{figure}[!h]
  \centering
  \includegraphics[width=0.6\textwidth]{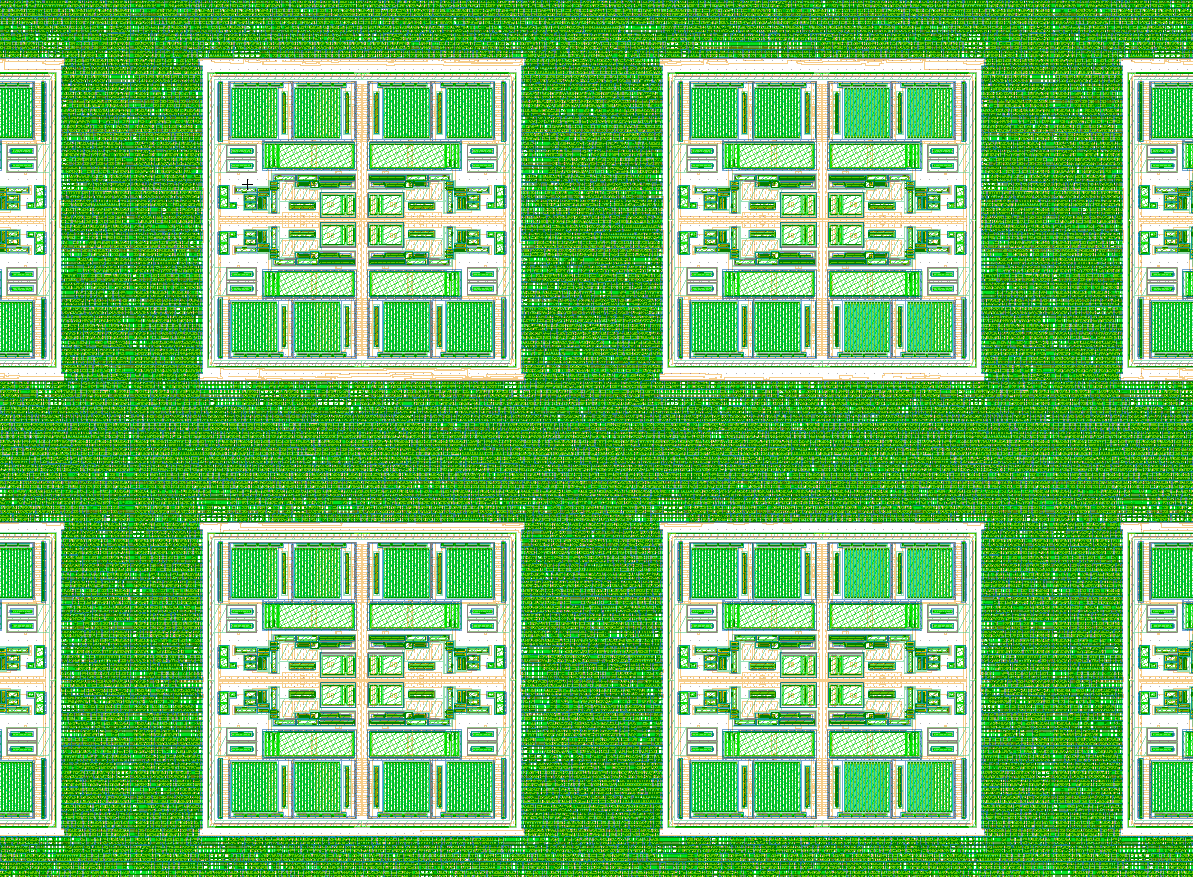}
  \caption{Layout detail of FE65-P2 showing analog islands surrounded by synthesized logic. Four whole islands can be seen in the center of the figure. Some layout layers have been suppressed for clarity.}
\label{fig:islands}
\end{figure}

The front ends are laid out in compact groups of four pixels or quads (also called analog islands) sharing power and bias distribution, as shown in Fig.~\ref{fig:islands}.
The analog islands are surrounded on all sides by synthesized logic, isolated from the substrate by a deep N-well. Each island is in its own, separate deep N-well in the double isolated chip, or directly on the substrate in the digital isolated chip.
Configuration bits for front end tuning and function selection are stored in the synthesized digital logic and supplied to each front end quad as static CMOS levels.

We measured two each double isolated and digital isolated FE65-P2 chips. Each bare die was mounted on its own passive test card, which could be connected to an active interface board to power, computer and instruments. A useful FE65-P2 feature for this study is the ability to disable all digital activity in the pixel matrix and still have access to the discriminated analog pixel outputs, one at a time, via a dedicated global OR hit output. We refer to this mode as ``digital off". This permits observing an ideal critical threshold baseline for one pixel at a time.

\section{Method}
All measurements were done with the same interface card and identical setup and control software, so that the only difference between the double isolated and digital isolated chip measurements was the chip itself. We take as a figure of merit of the isolation performance the observed single pixel noise increase caused by selected aggressor signals applied to the digital domain only. Thus any observed noise increase from aggressor off to aggressor on is the result of these aggressors coupling from the digital to the analog domain.

\begin{figure}[!h]
  \centering
  \subfigure[]{\includegraphics[width=0.49\textwidth]{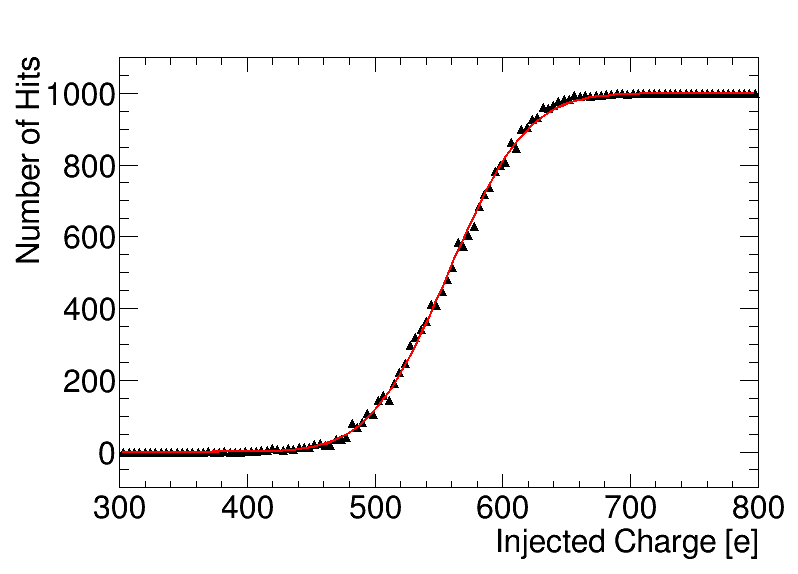}}
  \subfigure[]{\includegraphics[width=0.49\textwidth]{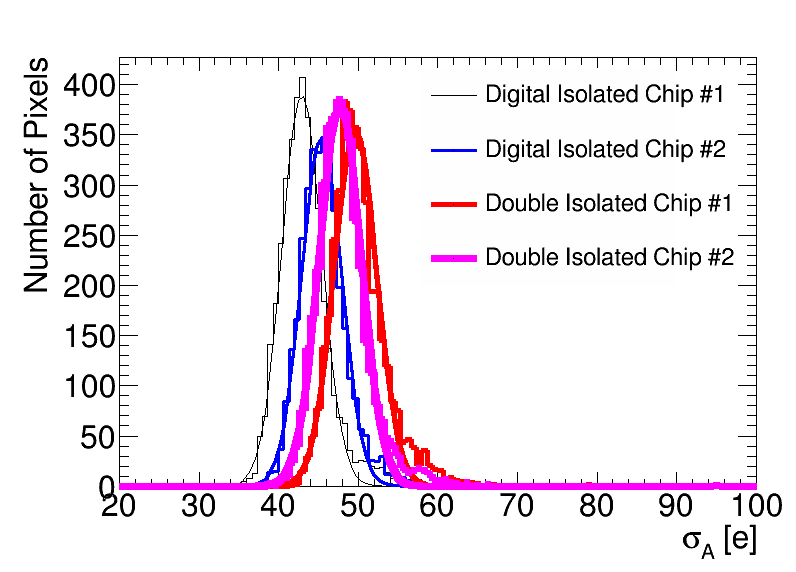}}
  \caption{(a) Illustration of S-curve obtained by counting number of hits with changing the injected charge. (b) Intrinsic analog front end noise obtained with digital off in each pixel of all measured chips (colour online).}
\label{fig:sigmaA}
\end{figure}

The intrinsic analog front end noise ($\sigma_\mathrm{A}$) is obtained with digital off by fitting an S-curve to the response counts vs. injected charge of each pixel discriminator output, as illustrated in Fig.~\ref{fig:sigmaA} (a). Fig.~\ref{fig:sigmaA} (b) shows the pixel-to-pixel distributions of the width of the S-curve ($\sigma_\mathrm{A}$) for all measured chips. The difference of the noise distribution between two different chips of the same type is comparable with the dispersion observed with ten FE-I4B wafers in Ref.~\cite{ref:dispersion}. 
The injection of test charges potentially introduces a small bias. Therefore, we use the relative measurement of the noise increase, which is insensitive to the absolute noise value measured here.

To measure the noise differences in a more sensitive and unbiased way, we do not use charge injection and S-curve reconstruction, but instead measure pixel noise occupancy (NOCC) as a function of the threshold. The NOCC is defined here as the number of noise hits observed in one pixel and in a time interval of one second, which is the same as the mean pixel firing frequency. Thus a NOCC of 10$^{-1}$ means that a pixel free-fires at an average rate of 0.1\,Hz. At a high value of threshold there will be zero NOCC. As the threshold is gradually decreased, the NOCC begins to rise exponentially~\cite{ref:spieler}. We plot the NOCC vs. threshold on a log scale and find the intercept of a line fit (on the log scale) with a ``floor'' occupancy of $0.1$. We call this intercept the critical threshold.

The critical threshold is a very sensitive measure of noise. We assume that the noise is Gaussian and that at
zero threshold the NOCC is of order $10^{6}$. This is determined by the response time of the Front End (including discriminator recovery time). 
Two distinct noise hits must be a few 100\,ns apart or they would be merged into one. 
The critical threshold of 0.1 is about 5$\sigma$ away from zero. Therefore, if we observe a shift of the critical theshold by an amount $x$
in response to an aggressor signal, this represents a change in noise $\sigma$ of $x/5$. Thus we can
detect smaller noise changes than possible by fitting an S-curve. Additionally, because the NOCC
measurement does not use any charge injection, it does not introduce perturbations that can bias
the measurement.

We use two ways to measure NOCC: (a) direct transition frequency measurement of the hit output of a single
pixel at a time, using an edge triggered frequency counter, and (b) counting recorded hits using the normal
triggered readout of the pixel matrix, by enabling a large number of pixels at a time and repeatedly triggering the 
chip readout.

\section{Single pixel measurements without noise injection: method (a)}
Method (a) makes use of the dedicated global OR hit output. 
With the clock disabled the chip is effectively a pure analog circuit and any noise observed is the intrinsic front end noise. When the clock is enabled and the digital circuitry is active, the possibility arises (due to less than perfect isolation) for digital noise to couple to the analog domain. This would show up as an increase in the measured noise.

\begin{figure}[!h]
  \centering
   \subfigure[]{\includegraphics[width=0.49\textwidth]{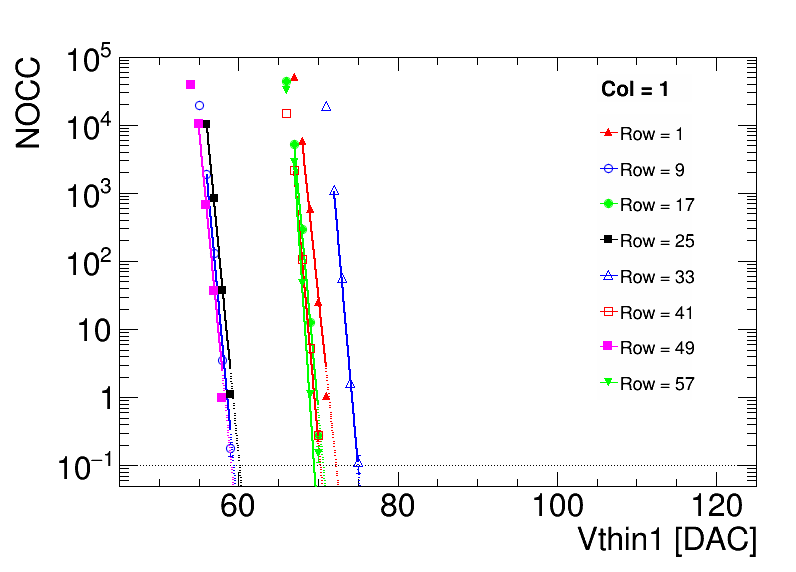}}
   \subfigure[]{\includegraphics[width=0.49\textwidth]{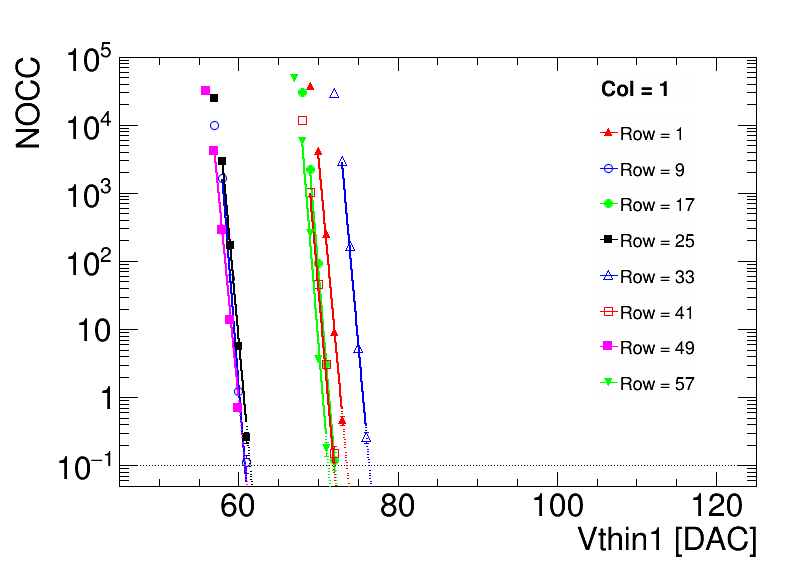}}
   \caption{Single pixel NOCC vs. threshold data for a few selected pixels with digital off (a) or on (b) in the first pixel column of the double isolated chip, obtained by counting recorded hits pixel by pixel using the global OR hit ouput of the discriminated analog (colour online).}
\label{fig:hitor1}
\end{figure}

Each pixel has a threshold adjustment called trim, but this has a small dynamic range. Additionally, there is a large dynamic range adjustment common to all pixels called global threshold, given by the Vthin1 and Vthin2 voltages in Fig.~\ref{fig:analog}. 
The trim threshold voltage in the pixel being studied is always set to the lowest value, to make the pixel have the lowest threshold in the chip for any given setting of the global threshold voltage.  
The trim threshold voltage of all other pixels are set to the highest value, in order to prevent them from becoming noisy as the global threshold is decreased.  
The threshold is changed by only changing Vthin1. Fig.~\ref{fig:hitor1} shows the single pixel NOCC vs. Vthin1 with digital off or on, for a few selected pixels on one of our tested chips, along with straight line fits intersecting the 0.1\,Hz horizontal line to extract the intercept which is called critical Vthin1. 

\begin{figure}[!h]
  \centering
  \includegraphics[width=0.6\textwidth]{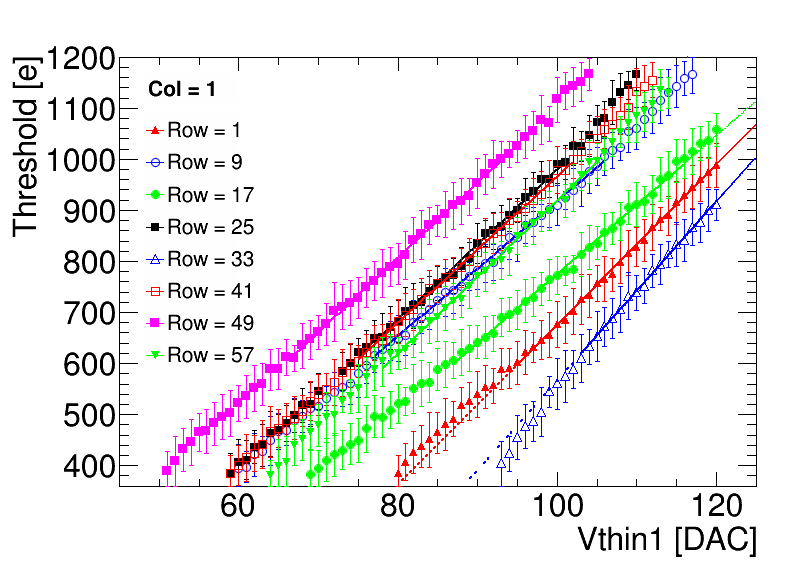}
  \caption{Threshold vs. Vthin1 for a few selected pixels in one of the double isolated chips (colour online).} 
\label{fig:thresholdVSvthin1dac}
\end{figure}

Fig.~\ref{fig:thresholdVSvthin1dac} illustrates the threshold in electrons as a function of Vthin1.
The threshold vs. Vthin1 distribution is fit with a linear function in the threshold range of 600\,[e] to 1000\,[e]. 
The critical threshold is calculated with critical Vthin1 and the relationship between threshold and Vthin1. 

\begin{figure}[!h]
  \centering
  \includegraphics[width=0.6\textwidth]{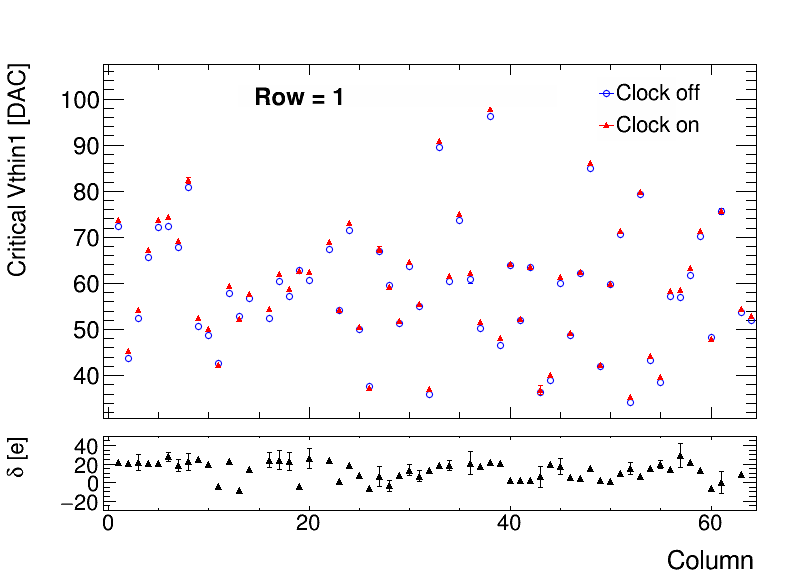}
  \caption{The change of the critical threshold when digital is on relative to digital off in the pixels of the first row of the double isolated chip.}
\label{fig:hitor-clock}
\end{figure}
Fig.~\ref{fig:hitor-clock} shows the change of critical threshold (referred as $\delta$[e]) when digital is on relative to digital off,  which indicates the aggressor effect of clocking the chip logic relative to the clock off baseline.

Based on the intrinsic front end noise width ($\sigma_\mathrm{A}$) and the width of the noise which spreads from digital to analog ($\sigma_\mathrm{D}$), the total noise obtained in analog with digital on would be $\sigma_\mathrm{A\bigotimes D}=\sqrt{\sigma_\mathrm{A}^{2}+\sigma_\mathrm{D}^2}$. If we assume the critical threshold is at the 5$\sigma$ of the noise distribution, then we have $\delta=5\cdot\sigma_\mathrm{A\bigotimes D}-5\cdot\sigma_\mathrm{A}$. Therefore, $\sigma_\mathrm{D}$ could be calculated with $\sqrt{(\delta/5+\sigma_\mathrm{A})^{2}-\sigma_\mathrm{A}^{2}}$.

\begin{figure}[!h]
  \centering
  \includegraphics[width=0.6\textwidth]{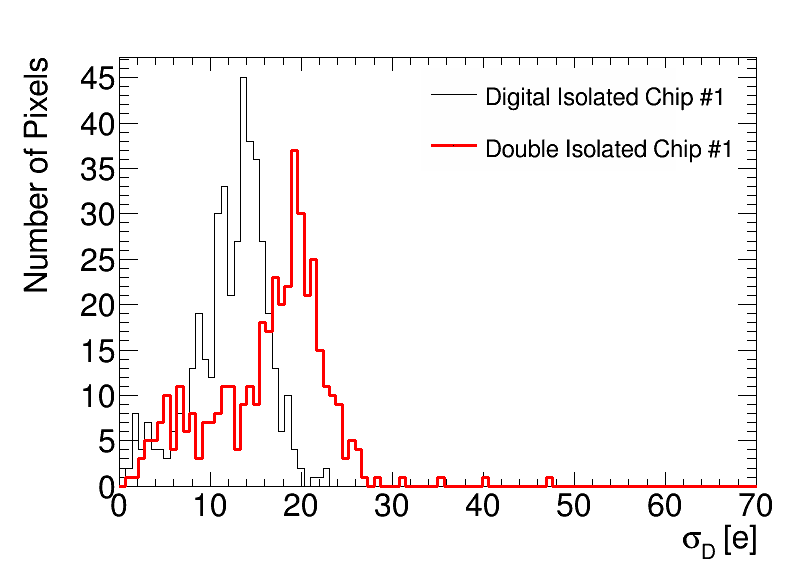}
  \caption{Intrinsic $\sigma_\mathrm{D}$ obtained without noise injection using method (a) (colour online).}
\label{fig:hitor-readout}
\end{figure}

Fig.~\ref{fig:hitor-readout} shows the aggressor effect of the pixel readout measured using method (a). While many pixels are enabled and read out, only one pixel at a time can be measured. The result is shown for a few measured pixels.
$\sigma_\mathrm{D}$ distributions shown in Fig.~\ref{fig:hitor-readout} imply that the digital isolation is better. However we do not control the characteristics of the noise injected by digital activity. To get some kind of noise transfer function from digital to analog we need to inject noise directly into the digital domaini, as presented in section 4.

\section{Pixel matrix measurements: method (b)}
\begin{figure}[!h]
  \centering
  \includegraphics[width=0.6\textwidth]{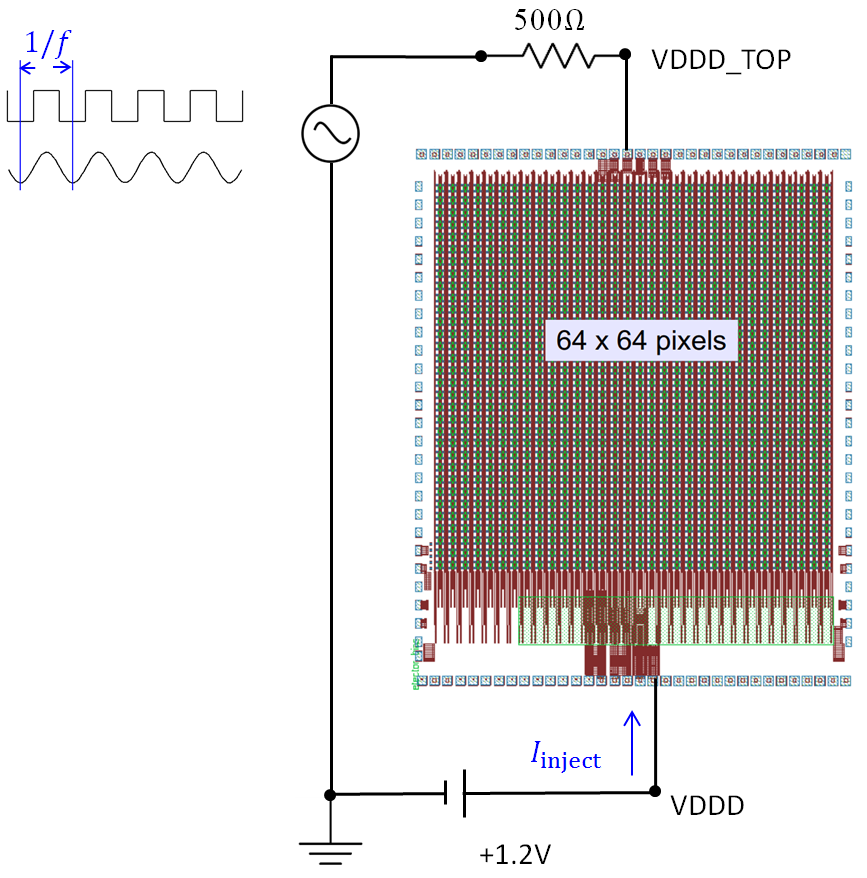}
  \caption{Schematic of current injection ($I_\mathrm{inject}$) to digital. The digital power supply to VDDD is kept at 1.2\,V. The highest voltage of the A/C signal (with frequency $f$) to VDDD\_TOP is always 1.2\,V, while the lowest voltage changes among different measurements.} 
\label{fig:injection-schematic}
\end{figure}
Method (b) uses the normal triggered readout of the pixel matrix by enabling a large number of pixels at a time. The digital is always on. The method exploits the FE65-P2 feature of power monitoring pads at the chip top (while the power is supplied from the chip bottom). Thus, in FE65-P2 we are able to run an externally controlled current through the digital power distribution network, independently from the chip current consumption. This current is our noise injection method. The injection method is shown schematically in Fig.~\ref{fig:injection-schematic}. The digital power supply at the chip bottom (VDDD) is kept at 1.2\,V as usual, while an A/C signal to the monitoring pad at the chip top (VDDD\_TOP) is used to draw an A/C current across the external load resistor. The highest voltage of the A/C signal (with frequency $f$) is always 1.2\,V, while the lowest voltage changes among different measurements to produce different amplitude of current ($I_\mathrm{inject}$).  The analog power supply from the chip bottom is kept at 1.2\,V. Analog and digital grounds are separated on the chip all the way to the bonding pads.

\begin{figure}[h]
  \centering
   \subfigure[]{\includegraphics[width=0.48\textwidth]{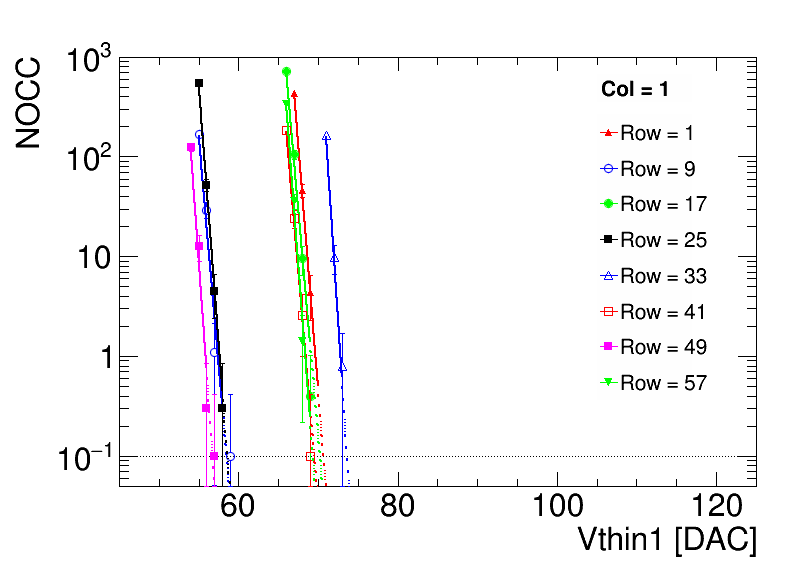}}
   \subfigure[]{\includegraphics[width=0.48\textwidth]{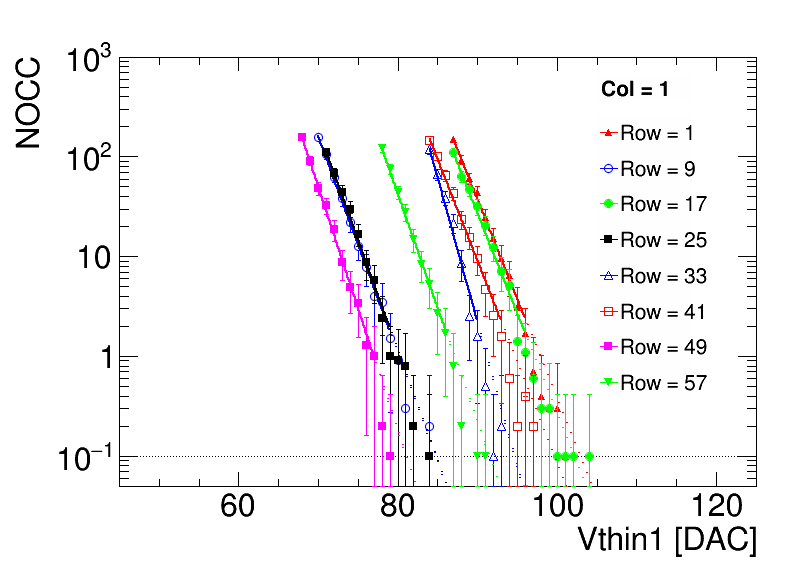}}
   \caption{Single pixel NOCC vs. threshold data for selected pixels before (a) and after (b) injecting 4.8\,mA current by an square wave with $f=1$\,MHz, obtained by counting recorded hits using the normal trigged readout of the pixel matrix (colour online).}
\label{fig:WF_hitVSvthreshold}
\end{figure}
The first measurement is based on square wave, the frequency of which is 1\,MHz. The injected current amplitude was 4.8\,mA, which corresponds to 20.87\% of the internal digital current consumption. Fig.~\ref{fig:WF_hitVSvthreshold} shows the single pixel NOCC as a function of the threshold for a few selected pixels on one of our tested chips, along with straight line fits intersecting the 0.1\,Hz horizontal line to extract the critical threshold. The critical threshold increased significantly due to the current changes in pixel columns. The change of the critical threshold when injecting the current fluctuation is $\delta=5\cdot\sigma_\mathrm{A\bigotimes D'}-5\cdot\sigma_\mathrm{A\bigotimes D}$. $\sigma_\mathrm{A\bigotimes D'}$ ($\sigma_\mathrm{A\bigotimes D}$) is the width of the noise observed in analog with (without) injecting current when digital is on. $\sigma_\mathrm{A\bigotimes D}$ is obtained with digital on by fitting an S-curve to the response counts vs. injected charge of each pixel discriminator ouput. 
$\sigma_\mathrm{A\bigotimes D'}$ is calculated with $\delta/5+\sigma_\mathrm{A\bigotimes D}$. Therefore, the width of the noise spreading from digital to analog with injecting current and digital on, $\sigma_\mathrm{D'}$, could be calculated with $\sqrt{\sigma_\mathrm{A\bigotimes D'}^{2}-\sigma_\mathrm{A}^{2}}$.

\begin{figure}[!h]
  \centering
  \includegraphics[width=0.5\textwidth]{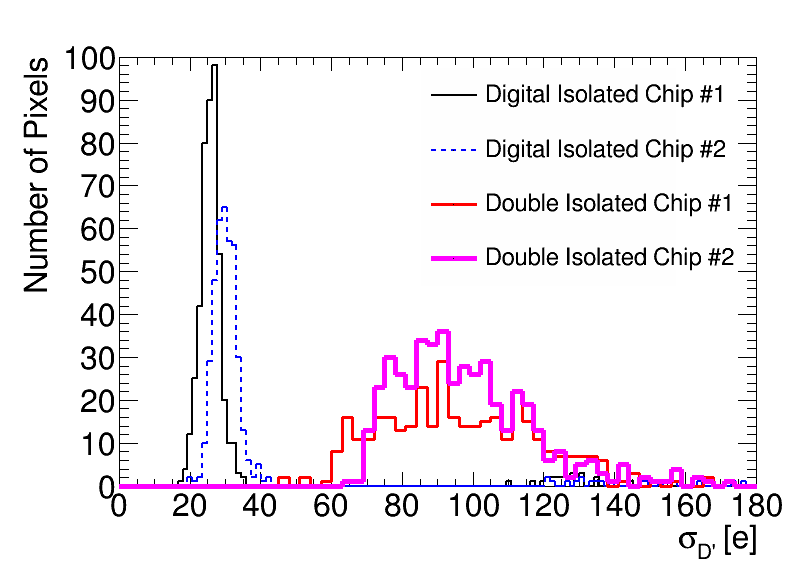}
  \caption{$\sigma_\mathrm{D'}$ obtained by injecting 4.8\,mA current (1MHz square wave) to digital (colour online).}
\label{fig:squareWF_delta_comp}
\end{figure}

Fig.~\ref{fig:squareWF_delta_comp} shows the $\sigma_\mathrm{D'}$ distribution when 4.8\,mA current (1\,MHz square wave) is injected into the digital power domain. This measurement with method (b) also indicates the digital isolation is better than double isolation.

\begin{figure}[!h]
  \centering
  \includegraphics[width=0.5\textwidth]{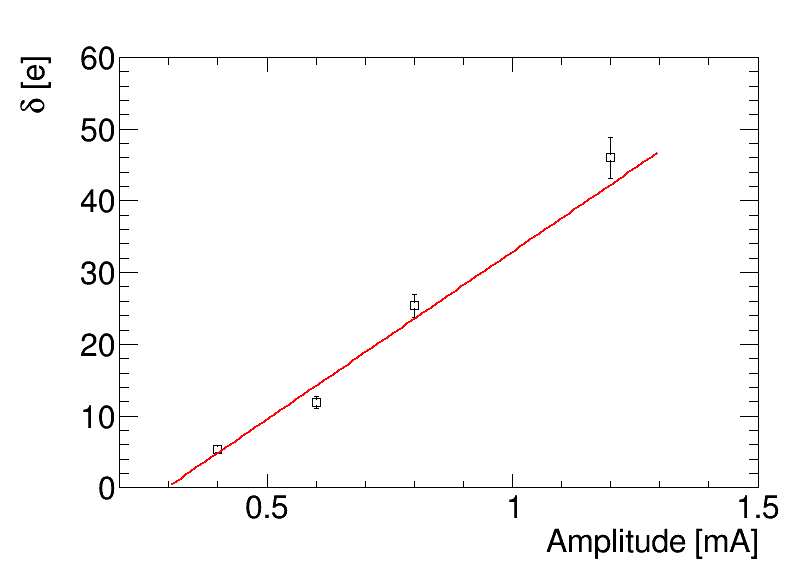}
  \caption{$\delta$ obtained by injecting different amplitude of current (18\,MHz sine wave) to digital in one of the double isolated chips.}
\label{fig:oldChip1_delta_vs_amp}
\end{figure}

\begin{figure}[!h]
  \centering
  \subfigure[]{\includegraphics[width=0.48\textwidth]{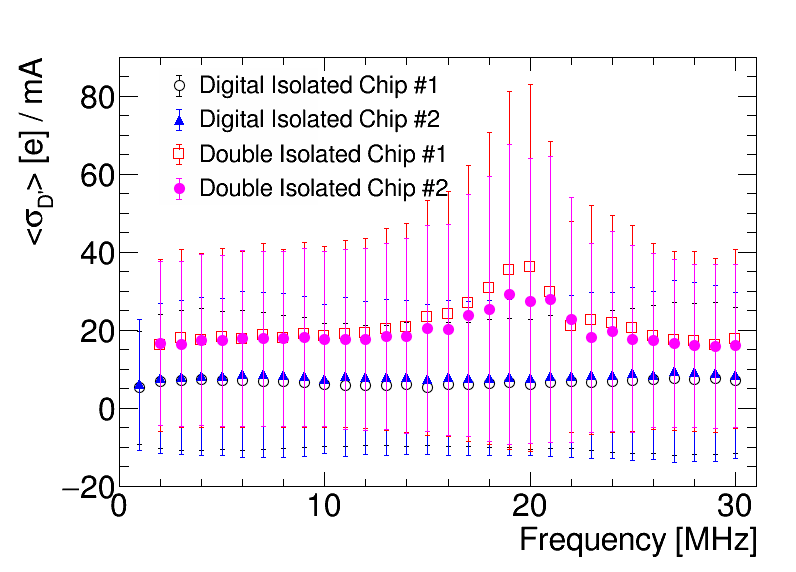}}
  \subfigure[]{\includegraphics[width=0.48\textwidth]{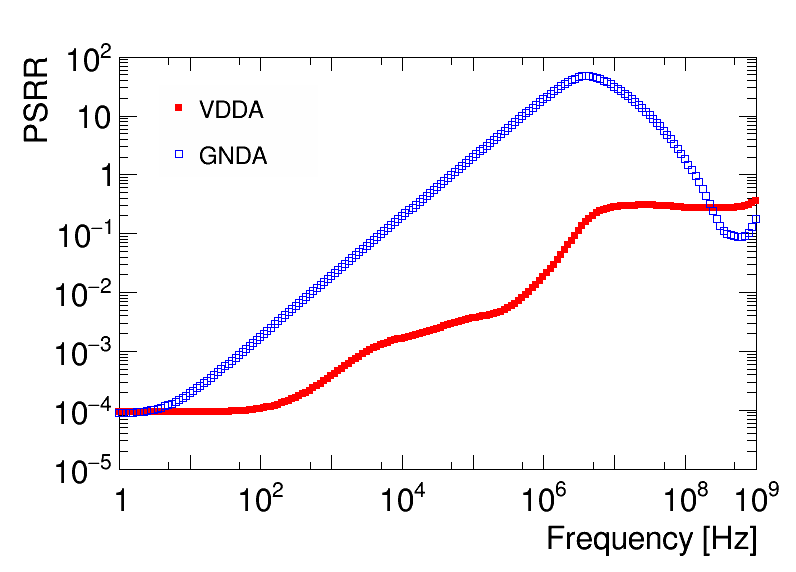}}
  \caption{(a) Average of $\sigma_\mathrm{D'}$ per mA vs the injected current frequency of sine wave in all of measured chips. (b) Simulated PSRR vs frequency for the power and ground rails of the front end in the double isolated chip.}
\label{fig:sineWF_delta}
\end{figure}

To study the frequency dependence of the noise coupling we use a sine wave instead of a square wave. 
Due to the observed difference in noise coupling, the double (digital) isolated chips are measured using a sine wave with 1.2\,mA (2.4\,mA) current injected to digital. Lower injection in the double isolated chip is necessary to avoid an excessive noise that can interfere with the chip operation.  
To compare the measurements, the change of critical threshold is divided by the amplitude of the injected current. 
Fig.~\ref{fig:oldChip1_delta_vs_amp} shows the linear dependence of the critical threshold change vs. the amplitude of the injected current. Each point (uncertainty) is the mean (statistical error in the mean) of the critial threshold change in all the pixels measured.  
Fig.~\ref{fig:sineWF_delta} (a) shows the average of $\sigma_\mathrm{D'}$ per mA among the measured pixels in each chip. The uncertainty of each measurement point is the root mean square of the $\sigma_\mathrm{D'}$ distribution. In comparison with the double isolated chips, the $\sigma_\mathrm{D'}$ as a function of the frequency is flatter and lower in the digital isolated chips. 
The similarity between the double isolated chip frequency response and the front end ground PSRR suggests that they are connected. 
Fig.~\ref{fig:sineWF_delta} (b) shows the simulated Power Supply Rejection Ratios (PSRR) for the power and ground rails of the front end design. As power and ground are common to all pixels, the simulation includes the full 64x64 pixel matrix with extracted parasitics, as well as estimated board parasitics and components added by hand. The gain peak for ground PSRR is a feature of the front end, as it is a single-ended ground-referenced design, but the exact frequency of the peak depends on the full distribution network and parasitics. While the peaks in Fig.~\ref{fig:sineWF_delta} (a) and (b) are not exactly at the same frequency, the comparison suggests that the digital noise coupling in the double-isolated chip is due to noise injection from the digital power network to the front end ground, rather than through the isolated substrate. This is plausible, because in the double isolated chip the ground well of each front end necessarily has a higher impedance to the common ground, as the substrate is not participating in the ground network. This local isolated well ground is easier to ``shake" than the massive substrate ground in the digital isolated chip. The noise in the digital metal power grid can directly couple to the analog ground this way. So Fig.~\ref{fig:sineWF_delta} suggests that metal to well coupling is the reason the double isolated chip is worse than the digital isolated chip. Removing the isolated analog well reduces this coupling. 

\section{Conclusion}
Two methods have been applied to measure the isolation between analog and digital. 
Our measurements indicate that the digital isolation which isolates only digital circuits from the substrate is better than the double isolation which isolates both analog and digital circuits from the substrate. Nevertheless, both isolation strategies perform well for normal operation (Fig.~\ref{fig:hitor-readout}): attempts to inject voltage noise on the digital supply failed to produce any measurable effect, and only the current injection (method (b)) succeeded.
The reason why the digital isolation is better than the double isolation might be that the noise is coupling through the metal stack rather through the substrate. In this case the double isolation has a higher impedance analog ground that is easier to shake by noise coming from the metal stack.

\section{Acknowledgements}
This work was supported by the U.S.~Department of Energy, Office of Science under contract DE-AC02-05CH11231. The chip fabrication was supported by the RD53 Collaboration.

\section{Appendix}
\begin{figure}[!h]
  \centering
  \includegraphics[width=0.5\textwidth]{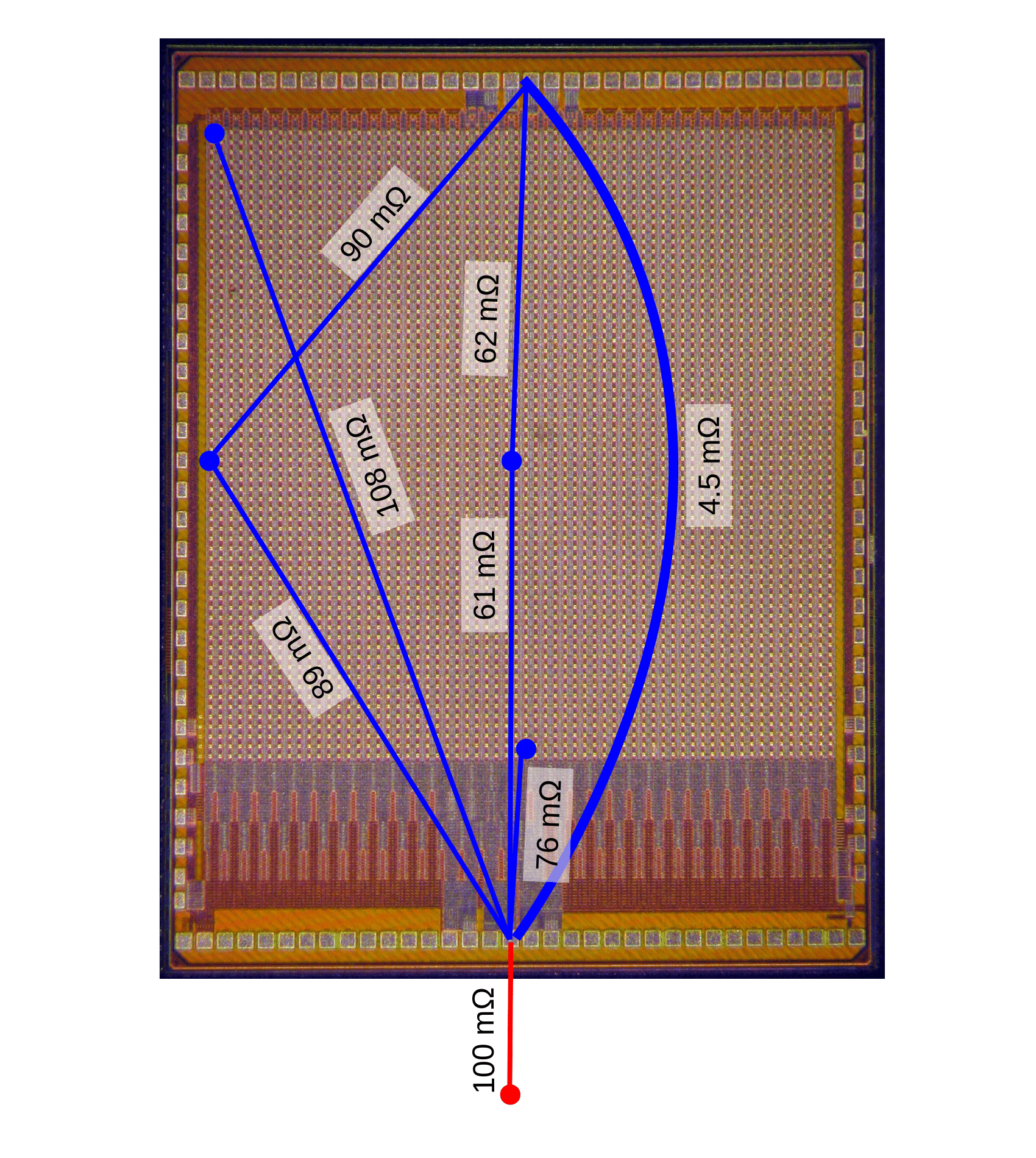}
  \caption{Point to point resistances for select extracted from the layout using parasitic extraction tools.}
\label{fig:resis}
\end{figure}

To evaluate the expected uniformity in the pixel matrix, the point to point resistances for select  were extracted from the layout using parasitic extraction tools. These are shown in Fig.~\ref{fig:resis}. The current flowing into a given digital pixel region of the layout is of order 0.1\% of the total matrix current, while the current flowing through the input wire bond is the total current. The results of Fig.~\ref{fig:sineWF_delta} indicate that the pixel matrix voltage will be very uniform compared to the average value fluctuations. For a total current variation $\Delta$I, the average voltage will fluctuate mainly due to the wire bond resistance by 0.1$\Delta$I, while the difference between  the most different points in the matrix will be 4$\times 10^{-4}\Delta$I $-$ a 0.4\% non-uniformity. This is consistent with our failure to observe any pixel position dependence in the digital to analog noise coupling.


\section{References}
\begin{thebibliography}{99}

\bibitem{ref:FE65-P2}
M Garcia-Sciveres et al., Results of FE65-P2 Pixel Readout Test Chip for High Luminosity LHC upgrades, Proc. 38th Int. Conf. on High Energy Physics (2016) p272.

\bibitem{ref:isolation}
The RD53 collaboration, Recent progress of RD53 Collaboration towards next generation Pixel Read-Out Chip for HL-LHC, JINST 11 (2016) C12058.

\bibitem{ref:dispersion}
M Backhaus, Characterization of the FE-I4B pixel readout chip production run for the ATLAS Insertable B-layer upgrade, JINST 8 (2013) C03013.

\bibitem{ref:spieler}
S Helmuth, Semiconductor Detector Systems, Semiconductor Science and Technology, Oxford University Press, 2005.  


\end{thebibliography}
\end{document}